\documentclass[10pt]{article}

\usepackage[margin=0.9in]{geometry}
\usepackage{
  amsmath, amssymb, amsfonts, mathtools, bm,
  graphicx, float, caption, booktabs, multirow, longtable,
  enumitem, cite, hyperref, xcolor, tikz, parskip, titlesec
}

\hypersetup{
    colorlinks = true,
    linkcolor = blue,
    urlcolor = blue,
    citecolor = red
}

\definecolor{sectioncolor}{HTML}{1A237E}
\definecolor{subsectioncolor}{HTML}{1565C0}
\definecolor{subsubsectioncolor}{HTML}{00796B}

\titleformat{\section}
  {\color{sectioncolor}\normalfont\Large\bfseries}
  {\thesection}{1em}{}

\titleformat{\subsection}
  {\color{subsectioncolor}\normalfont\large\bfseries}
  {\thesubsection}{1em}{}

\titleformat{\subsubsection}
  {\color{subsubsectioncolor}\normalfont\normalsize\bfseries}
  {\thesubsubsection}{1em}{}

\renewcommand{\arraystretch}{1.2}
\setlist[itemize]{leftmargin=1.5em}

\usepackage{amsmath, amssymb, amsfonts,footmisc}
\usepackage{graphicx,xcolor,fdsymbol,fancyhdr}
\usepackage[dvipsnames]{xcolor}
\usepackage{booktabs, rotating,multirow,lscape}
\usepackage{hyperref}
\usepackage{algorithm2e}
\usepackage{caption}
\usepackage{cite}
\newcommand{\club}[0]{\clubsuit} 
\newcommand{\diamondc}{\color{red}\vardiamondsuit\color{black}} 
\newcommand{\spade}[0]{\spadesuit} 
\newcommand{\heart}[0]{\color{red}\varheartsuit\color{black}}
\title{\textbf{Adapting Skill Ratings to Luck-Based Hidden-Information Games}}

\author{
\begin{tabular}{@{}c@{\hspace{1.5cm}}c@{}}
\textbf{Avirup Chakraborty} & \textbf{Shirsa Maitra} \\
\small Indian Statistical Institute & \small Computer Science Engineering \\
\small Kolkata, India & \small Heritage Institute of Technology, India \\[0.4cm]
\textbf{Tathagata Banerjee} & \textbf{Diganta Mukherjee} \\
\small Department of Statistics \& Data Science & \small Sampling and Official Statistics Unit \\
\small National University of Singapore & \small Indian Statistical Institute, Kolkata \\[0.4cm]
\end{tabular}
\\[0.2cm]
\begin{tabular}{@{}c@{}}
\textbf{Tridib Mukherjee} \\
\small Chief Data Scientist and AI Officer\\
\small IDfy, India\\
\end{tabular}
}

\begin{document}
\date{}
\maketitle
\begin{abstract}
Rating systems play a crucial role in evaluating player skill across competitive environments. The Elo rating system, originally designed for deterministic and information-complete games such as chess, has been widely adopted and modified in various domains. However, the traditional Elo rating system only considers game outcomes for rating calculation and assumes uniform initial states across players. This raises important methodological challenges in skill modelling for popular partially randomized incomplete-information games such as Rummy. In this paper, we examine the limitations of conventional Elo ratings when applied to luck-driven environments and propose a modified Elo framework specifically tailored for Rummy. Our approach incorporates score-based performance metrics and explicitly models the influence of initial hand quality to disentangle skill from luck. Through extensive simulations involving 270,000 games across six strategies of varying sophistication, we demonstrate that our proposed system achieves stable convergence, superior discriminative power, and enhanced predictive accuracy compared to traditional Elo formulations. The framework maintains computational simplicity while effectively capturing the interplay of skill, strategy, and randomness, with broad applicability to other stochastic competitive environments.
\end{abstract}


\section{Introduction}
Player skill evaluation is a fundamental component of competitive gaming ecosystems, guiding matchmaking, player progression, and strategic analysis. Traditional rating systems, such as the Elo \cite{elo1978}, Glicko \cite{glickman1999}, and TrueSkill \cite{herbrich2006} frameworks, have proven effective in deterministic or near-deterministic games like chess and Go, where outcomes primarily reflect player skill. However, their direct application to luck-influenced, incomplete-information games such as Rummy, Poker, or Bridge introduces notable limitations. In such games, outcomes are not solely determined by player decisions but also by stochastic elements like initial card distributions and variable round dynamics. Consequently, standard Elo-type formulations tend to overestimate short-term fluctuations and underestimate persistent skill differences.

The Rummy environment presents a unique case within this class. It combines probabilistic uncertainty (due to card draws and opponent hands) with strategic depth (melding, discarding, bluffing), leading to outcomes that depend on both tactical skill and random events. This dual dependence makes player-skill inference inherently noisy, requiring rating mechanisms that can disentangle luck from skill and stabilize ratings over repeated play. Existing systems often fail to achieve this balance. Rapid volatility in ratings discourages consistent player engagement, while slow convergence obscures meaningful skill differentiation. Our study will be focused solely on Rummy. 

Recent advances in rating systems, such as TrueSkill Through Time \cite{dangauthier2007} and TrueSkill 2 \cite{minka2018}, have focused on improving dynamic updating and parameter learning; however, these models remain primarily suited to deterministic or team-based settings. In contrast, Edelkamp et al (2021) \cite{edelkamp2021} proposed an Elo-based framework for Skat, a game characterized by significant random variation, demonstrating the necessity of correcting standard Elo formulations to account for chance effects. Building on the broader discussion of skill versus luck, Banerjee et al. (2021) \cite{Tathagatada} introduced a data-driven statistical framework to quantify the relative influence of skill and chance across popular games, revealing systematic differences among Chess, Rummy, Ludo, and Teen Patti. Together, these studies underscore that rating systems for games involving both randomness and hidden information, such as Rummy, must explicitly incorporate chance-adjustment mechanisms to produce fair and stable estimates of player skill.

This paper introduces a modified Elo formulation tailored for Rummy, which can be suitably adjusted to other games played in randomized environments. The method is designed to incorporate luck-adjusted weighting and match-context normalization while maintaining computational simplicity. Further, our method is also adaptable to deterministic games as special cases of parameter choices. Our approach aims to (i) model the probabilistic uncertainty intrinsic to Rummy’s deal structure, (ii) reduce the inflation and deflation effects that arise from streaks of fortunate or unfortunate draws, and (iii) preserve interpretability comparable to the classical Elo scale. The proposed method is evaluated through extensive simulations across varying skill distributions, comparing its stability, convergence rate, and predictive reliability against conventional Elo \cite{elo1978}, Glicko \cite{glickman1999}, and TrueSkill \cite{herbrich2006} frameworks. 

Section~\ref{sec:ratings} outlines the methodological framework, reviewing standard rating systems and highlighting their limitations in games influenced by chance before introducing our proposed formulation. Section~\ref{sec:method} details the proposed methodology and the corresponding parameter estimation procedures. Section~\ref{sec:simulation} describes the simulation design and presents the empirical findings. Section~\ref{sec:discussion} interprets the results, emphasizing how the proposed rating system more effectively captures underlying skill in stochastic environments compared to existing approaches. Finally, Section~\ref{sec:conclusion} concludes with potential extensions and practical applications, particularly for online Rummy platforms.

\section{Different Existing Rating Systems}
\label{sec:ratings}
This section provides an overview of several classical rating systems that have been widely used to evaluate player skill.

\begin{itemize}

    \item \textbf{Elo-Based Rating Systems:}  
    Elo-based systems model player skill as a latent variable that is iteratively updated based on game outcomes.  
    In the \textit{classical Elo} formulation for a two-player match between players $i$ and $j$, the expected outcome for player $i$ is modeled using a logistic link:
    \[
    \alpha_i = \frac{1}{1 + 10^{(R_j - R_i)/400}},
    \]
    and the rating update rule is
    \[
    R_i' = R_i + K (S_i - \alpha_i),
    \]
    where $S_i \in \{0, 0.5, 1\}$ denotes the observed outcome (loss, draw, or win), and $K$ is a fixed sensitivity parameter. Extensions of the classical model address uncertainty and adaptive learning rates. In \textit{Bayesian Elo}, each player's rating is treated as a Gaussian random variable, $R_i \sim \mathcal{N}(\mu_i, \sigma_i^2)$, allowing probabilistic updates. \textit{Elo-MMR} further introduces a player-specific sensitivity factor $K_i = f(\sigma_i, \sigma_j, t)$ that adapts to rating deviation and recency, enabling faster updates for uncertain players and greater stability for consistent ones.
    
    \vspace{0.2cm}
    \item \textbf{Glicko Systems:}  
    While the Elo system assumes a fixed confidence in player skill, it fails to account for uncertainty arising from limited or inconsistent play. 
    The \textit{Glicko} system addresses this by introducing a \textit{rating deviation (RD)} that quantifies uncertainty in a player's true strength. 
    A higher RD indicates that a player's current rating is less reliable, typical for newcomers or inactive players, while frequent and consistent play reduces RD, reflecting growing confidence in the estimated skill. 
    From the Elo formula, $K$ is scaled by the rating deviation (larger for uncertain players). 
    \textit{Glicko-2} extends this framework by adding a \textit{volatility parameter}  to capture fluctuations in player performance over time, enabling faster correction for unstable or streak-prone players. 

    \vspace{0.2cm}
    \item \textbf{TrueSkill and TrueSkill-2:}  
    \textbf{TrueSkill and TrueSkill-2:}  
    TrueSkill, developed by Microsoft, extends beyond Elo and Glicko in a Bayesian approach, treating each player's skill as a normal distribution $p_i \sim N(\mu_i, \sigma_i^2)$ rather than a fixed point estimate. In a two-player match:
    \[
        P(i \text{ wins}) = \Phi\left(\frac{\mu_i - \mu_j}{\sqrt{2(\sigma_i^2 + \beta^2)}}\right),
    \]
    where $\beta$ represents random performance variation within games.  
    After each match, both $\mu_i$ and $\sigma_i$ are updated through approximate Bayesian inference, allowing the model to capture learning and uncertainty dynamically. Unlike Elo or Glicko, which assume scalar and independent ratings, TrueSkill supports multiplayer and team-based games by jointly inferring skills from all outcomes.  
    TrueSkill-2 further refines this by enabling multidimensional skill vectors and real-time tracking, though at the cost of higher computational complexity and less transparency compared to classical rating systems.
\end{itemize}

Despite their success in modeling competitive skill, these systems exhibit limited inferential accuracy in games where outcomes are substantially affected by chance. To overcome this limitation, the next section details our proposed model, designed to incorporate the role of luck in performance evaluation.

\section{Methodology}
\label{sec:method}
This section outlines the methodological framework underlying our proposed model. Our objective is to develop a rating mechanism that not only captures player skill but also adjusts for randomization inherent in the game of Rummy. For completeness, a detailed description of the two-player Rummy setup is provided in Appendix~\ref{sec:rules}. 
\subsection*{Challenges}
Before presenting the model, we highlight two key challenges that motivate our approach:
\begin{itemize}
\item \textbf{Performance Differentiation:}  
Rummy is an incomplete information game, where players independently form melds that may not be directly influenced by their opponents’ actions. The final scores of both players provide richer information than just the game outcome: a small score difference suggests comparable play, while a large difference indicates a dominant performance.

\item \textbf{Influence of Initial Hands:}  
Rummy involves substantial randomness, and outcomes can be heavily affected by the initial hand. 
A poor starting hand significantly reduces a player’s chance of winning, and ignoring this factor introduces bias into skill estimates.
\end{itemize}
\subsection*{Formulation}

We formulate our model in a stepwise manner as follows:

\begin{itemize}
    \item We begin with the classical Elo update rule:
    \[
    R_i' \leftarrow R_i + K (S_i - E_i),
    \]
    where $S_i \in \{0, 0.5, 1\}$ denotes the observed match outcome (loss, draw, or win), $K$ is the sensitivity parameter, \(R_i\) and \(R_i'\) are the pre- and post-match ratings, and \(E_i\) is the expected outcome for player \(i\).
    
    In games with high randomness or incomplete information, such as Rummy, the expected outcome \(\alpha_i\) (that is, the probability of winning) can vary substantially due to factors such as the initial hand and hidden opponent information.

    \item To address this limitation, we adopt a \textbf{score-based formulation} instead of an outcome-based one. In Rummy, we have already introduced the metrics \textbf{MinScore} and \textbf{MinDist} (Appendix~\ref{minscore}), which quantify the \emph{declarability} of a given hand. Modeling these scores enables us to capture the underlying gameplay dynamics of a player more effectively. Accordingly, we reformulate the rating update as
    \[
    R_i' \leftarrow R_i + K (A_i - B_i),
    \]
    where \(A_i\) denotes the observed (final) score of player \(i\), and \(B_i\) represents a benchmark score modeled for that player. Since these scoring system is inherently bounded, the difference \(|A_i - B_i|\) remains within a finite range, allowing \(K\) to be scaled appropriately to maintain numerical stability.
    
    It is important to note that in Rummy, a lower \textbf{MinScore} or \textbf{MinDist} value indicates a better hand, with a score of zero signifying a successful declaration. Therefore, the direction of the rating adjustment must be inverted, effectively rendering the sign of \(K\) negative.

    \item To maintain a \textit{zero-sum} structure analogous to the Elo system, we impose the condition
    \[
    (A_1 - B_1) + (A_2 - B_2) = 0 
    \quad \Longrightarrow \quad 
    B_1 + B_2 = A_1 + A_2 = A \text{ (say)}.
    \]
    This establishes a linear dependency between the two benchmark scores that must be jointly determined.  

    We use the term \textit{benchmark} instead of \textit{expected outcome}, since the benchmark scores \(B_i\) depend on the realized total \(A_1 + A_2\) observed after the match, unlike the probabilistic expectation in the Elo model, which is computed beforehand.

    \item To estimate the values of \(B_i\), we consider two key factors. The rating difference between the players, \(D_R = R_1 - R_2\), represents the pre-game skill bias, while the difference in their initial hand scores, \(D_H = H_1 - H_2\), captures the randomness arising from card distribution. Incorporating both allows us to reduce two types of bias in benchmark estimation.  

    Using a logistic-type scaling function, we define
    \[
    B_1 = \frac{A}{1 + 10^{-(\alpha\times D_R + \beta\times D_H)}}, \quad
    B_2 = A - B_1 = \frac{A \cdot 10^{-(\alpha\times D_R + \beta\times D_H)}}{1 + 10^{-(\alpha\times D_R + \beta\times D_H)}},
    \]
    where \(\alpha\) and \(\beta\) are model parameters to be estimated empirically. This formulation ensures that \(B_i \in [0, A]\) for both players.

    \item As mentioned earlier, the parameter \(K\) changes sign due to the inverse nature of the scoring metrics. Therefore, \(|K|\) represents the sensitivity parameter governing the magnitude of rating adjustments. Now, Players who have played fewer games should have ratings that respond more strongly to new outcomes, reflecting greater uncertainty about their true skill level. Therefore, each player \(i\) may be assigned an individual sensitivity factor \(K_i\) based on their number of games played \(n_i\). A simple formulation is given by:
    \[
    K_i =
    \begin{cases}
    k_1, & \text{if } n_i \leq p, \\
    k_2, & \text{if } p < n_i \leq q, \\
    k_3, & \text{if } n_i > q,
    \end{cases}
    \quad i = 1, 2.
    \]
    Here, \(|k_1| > |k_2| > |k_3|\), ensuring that as a player gains more experience (larger \(n_i\)), their rating becomes progressively less sensitive to individual match outcomes.

    \item Finally, we consider the case where player 1 wins (similarly for player 2), implying \(A_1 < A_2\). If \(A_1 - B_1 > 0\) happens, i.e.  the player won but did not play as good as the benchmark, to avoid penalizing the winner we modify the rating updates as follows:
    \[
    R'_1 \leftarrow R_1 + K_1 (A_1 - B_1) \, \mathbf{1}_{\{(A_1 - B_1)(W_1 - 0.5) \leq 0\}},
    \]
    \[
    R'_2 \leftarrow R_2 + K_2 (A_2 - B_2) \, \mathbf{1}_{\{(A_2 - B_2)(W_2 - 0.5) \leq 0\}},
    \]
    where \(W_i\) is the game outcome indicator (1 for a win and 0 for a loss), and \(\mathbf{1}_{\{\cdot\}}\) denotes the indicator function. This ensures that rating adjustments are applied only when the score-based deviation aligns with the actual game result.
\end{itemize}
Thus, we complete the formulation of our proposed model. We initialize each player with a rating of 1000 and subsequently update these ratings according to the specified rules. Next, we outline the procedure for estimating the model parameters.

\subsection*{Parameter Estimation}
We estimate the model parameters—player abilities \(\alpha\) and \(\beta\), and concentration parameters \(K_i\)—using the \textbf{MinScore} metric. This choice is motivated by MinScore's broader dynamic range (0 to 80, per Appendix~\ref{minscore}) compared to the more limited \textbf{MinDist}. The increased variation in MinScore values provides greater discrimination between player performances, yielding more sensitive parameter estimates.

\textbf{\textit{Choice of \(K\):}}

In our simulation study (Section~\ref{sec:simulation}), we assume that all players participate in approximately the same large number of games. Consequently, the individual \(K_i\) values have negligible impact on the long-run results, as they converge after a sufficient number of matches. Hence, we set \(K_1 = K_2 = K\).  

At the end of a game, one of the players typically achieves a declarable hand, implying \(A_i = 0\) for that player, while the maximum possible score is \(A_i = 80\). From the model specification, the maximum possible change in rating per game is therefore \(80|K|\). To restrict the maximum rating variation to 50 points (i.e., 5\% of the initial rating of 1000), we have
\[
80|K| = 50 \implies |K| = 0.625.
\]
Since \(K < 0\) by construction, we set \(K = -0.625\). Naturally, choosing a different threshold for the maximum allowable rating change would yield a different value of \(K\), thereby altering the resulting rating distribution.

\textbf{\textit{Choice of Parameter \(\alpha\):}}

Before estimating the constants \(\alpha\) and \(\beta\), it is essential to understand their interpretation within the model. The core relationship is given by:
\[
\frac{B_1}{B_2} = 10^{\alpha D_R + \beta D_H},
\]
where a lower benchmark score \(B_i\) indicates better performance.

To isolate the effect of the rating difference \(D_R\), we consider the scenario with a neutral hand advantage (\(\beta D_H = 0\)). The model then simplifies to:
\[
\frac{B_1}{B_2} = 10^{\alpha D_R}.
\]

If Player~1 has a higher initial rating (\(D_R > 0\)), we expect better performance, meaning \(B_1 < B_2\) and thus \(B_1/B_2 < 1\). For the right-hand side, \(10^{\alpha D_R}\), to be less than 1, the exponent \(\alpha D_R\) must be negative. Since \(D_R > 0\), this necessitates that \(\alpha\) must be a \textit{negative} constant.

The magnitude of \(\alpha\) controls the sensitivity of the model. A smaller absolute value \(|\alpha|\) means that a larger rating difference \(D_R\) is required to significantly affect the benchmark ratio \(B_1/B_2\). This leads to a system where larger rating differences are needed to distinguish between players, effectively increasing the spread of ratings across the population. However, if \(|\alpha|\) is too small, the ratings may diverge excessively, making the system unstable. Therefore, \(\alpha\) must be fixed to a carefully chosen value to ensure stable and meaningful ratings.

We calibrate \(\alpha\) by considering an extreme, yet plausible, in-game scenario. Let us assume:
\begin{itemize}
    \item \(D_R = 500\), indicating Player~1 is substantially stronger than Player~2.
    \item \(D_H = 0\), indicating both players have hands of identical quality.
    \item The resulting scores are \(B_1 = 2\) (near-optimal) and \(B_2 = 80\) (the maximum non-zero score in 13-card Rummy).
\end{itemize}

Substituting these values into the model yields:
\[
\frac{2}{80} = 10^{500\alpha} \quad \Rightarrow \quad \alpha = \frac{-\log_{10}(40)}{500} \approx -0.0032.
\]
We use this estimate of \(\alpha\) in the subsequent simulation study.

\textbf{\textit{Choice of Parameter \(\beta\):}}

For parameter \(\beta\), consider the scenario where \(D_R = 0\), indicating players of equal skill levels:

\[
\frac{B_1}{B_2} = 10^{\beta D_H}
\]
when \(D_H > 0\), Player~1 faces a hand disadvantage, and we expect Player~2 to perform better, i.e., \(B_1 > B_2\) and consequently \(B_1/B_2 > 1\). For the right-hand side \(10^{\beta D_H}\) to exceed 1, the exponent \(\beta D_H\) must be positive. Since \(D_H > 0\), this requires \(\beta\) to be a \textit{positive} constant.

\begin{figure}[H]
    \centering
    \includegraphics[width=0.8\linewidth]{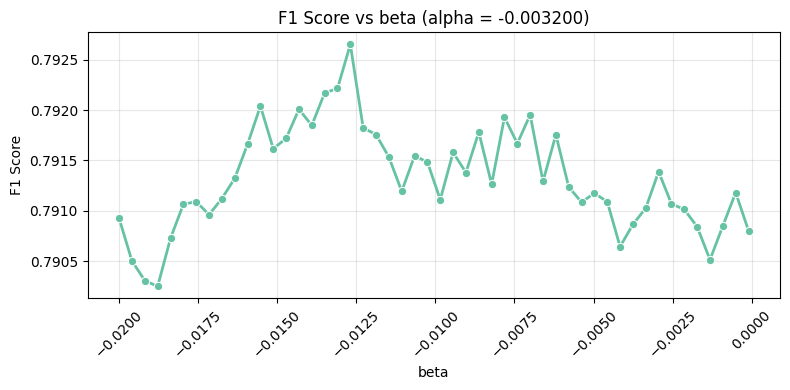}
    \caption{ }
    \label{fig:paramtune}
\end{figure}

We estimate \(\beta\) through a systematic tuning procedure with the following steps:

\begin{itemize}
    \item Fix \(\alpha = -0.0032\) as determined from our previous calibration
    \item Evaluate a range of \(\beta\) values by computing dynamic ratings on simulation data
    \item Partition the dataset into training and testing subsets
    \item Train a logistic regression model using only initial rating difference to predict match winners
    \item Assess model performance on the test set using F1 score as the primary metric
    \item Select the \(\beta\) value that maximizes predictive performance
\end{itemize}
To determine the optimal model configuration, we performed a hyperparameter search over a range of values for \(\beta\). The resulting F1 scores from this tuning process are plotted in Figure~\ref{fig:paramtune}. The analysis reveals a clear optimum at \(\beta = -0.012690\), which yields a maximum F1 score of 0.7927. This value of \(\beta\) optimally calibrates the model's update function, maximizing the harmonic mean between precision and recall.
The estimated values for \(K\), \(\alpha\), and \(\beta\) will be used to parameterize the model in the subsequent simulation study.

\section{ Simulation Study \& Results}
\label{sec:simulation}
In this section, we describe the simulation framework employed for the empirical evaluation of the competing strategies. Detailed descriptions of each strategy are provided in Appendix~\ref{strategy_descriptions}. These strategies are taken from \cite{Puruda}. Among the considered strategies, the Random and Defeat-Seeking agents serve as naive baselines, expected to perform poorly due to their lack of strategic sophistication. The MinDist and MinScore agents are based on the core distance and marginal-score metrics and therefore represent the fundamental benchmark strategies. The MinDist+MinScore Hybrid (Mindistscore) and Opponent-Aware MinDist (MindistOpp) extend the MinDist formulation by incorporating hybrid decision rules and opponent modelling, respectively. These enhanced variants are designed to exhibit stronger competitive performance.

The simulation study is structured as follows:
\begin{itemize}
\item We simulate 4,500 games for each ordered pair of strategies, resulting in \(6\times 5 =30\) directed matchups. Consequently, each strategy participates in \(4500\times5\times2 = 45,000\) games in total.
\item Each strategy plays an equal number of games as the first and second player to eliminate positional bias and ensure symmetry in competitive conditions.
\item The global sequence of games is randomised to avoid ordering effects and to ensure that rating updates evolve in a stochastic and unbiased manner.
\item For every match, we record the dynamic rating trajectories and rating increments, which will enable us to analyse rating stability and convergence over time.
\end{itemize}
Here are the results of the simulation,

\begin{figure}[H]
    \centering
    \includegraphics[width=0.8\linewidth]{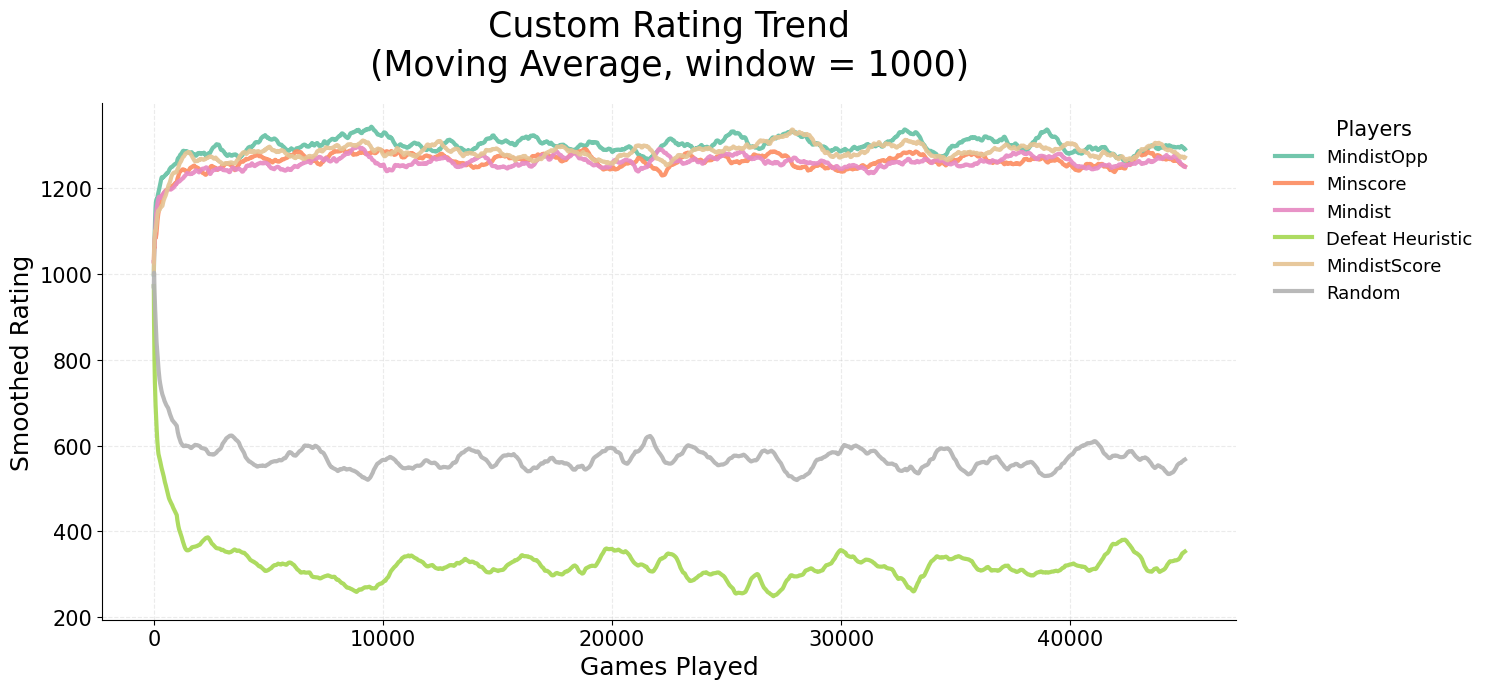}
    \caption{\textbf{Rating evolution over simulated games (Custom Elo)}}
    \label{fig:elo1}
\end{figure}

Figure~\ref{fig:elo1} illustrates how Elo ratings evolve throughout the simulation.

The following observations emerge:

\begin{itemize}
    \item Ratings begin to stabilise around 5,000 games, indicating convergence toward a consistent skill hierarchy and clear separation among strategies.
    
\end{itemize}

\begin{figure}[h]
    \centering
    \includegraphics[width=0.8\linewidth]{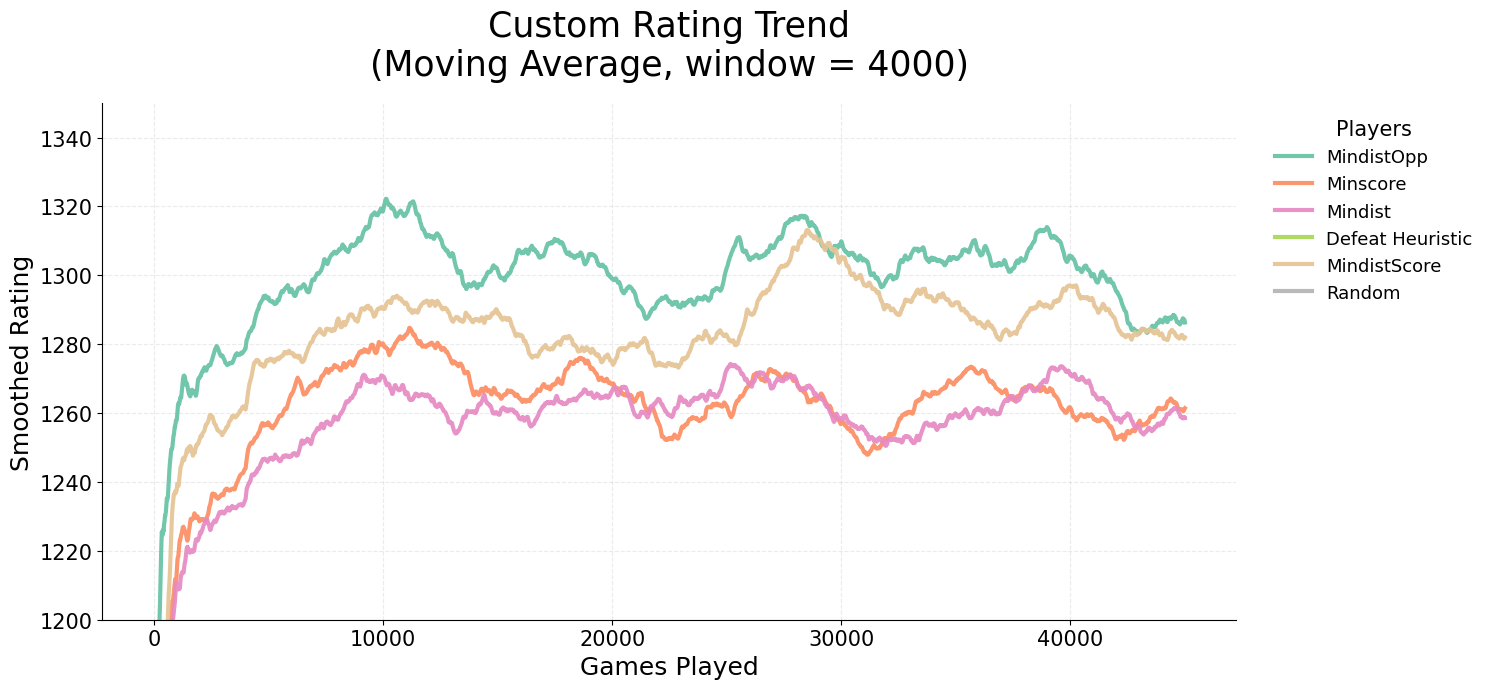}
    \caption{\textbf{Zoomed view of rating evolution for top strategies (Custom Elo)}}
    \label{fig:elo2}
\end{figure}

\begin{itemize}
    \item As anticipated, the naive strategies \textit{Defeat Heuristic} and \textit{Random} perform the worst. Their final ratings fall below the neutral Elo baseline of 1000, which confirms that they lose consistently across repeated play.

    \item Although a moving-average smoother is applied, small fluctuations remain. These can be attributed to heterogeneous matchups, where temporary pairings with stronger or weaker opponents cause short-term rating variation.
    
    \item Figure~\ref{fig:elo2} highlights the performance among the stronger strategies. The hybrid strategies \textit{MindistOpp} and \textit{MindistScore} consistently outperform the pure heuristics \textit{Mindist} and \textit{Minscore}, supporting the conclusion that combining distance and score information yields superior decision making capability.
\end{itemize}

Table~\ref{tab:rating_summary} reports summary statistics (mean, standard deviation, and coefficient of variation) for the rating paths. The results clearly demonstrate meaningful performance separation, with hybrid strategies achieving both higher and more stable ratings over time.

\begin{table}[H]
\centering
\renewcommand{\arraystretch}{1.25}
\setlength{\tabcolsep}{8pt}

\begin{tabular}{lrrr}
\toprule
\textbf{Strategy} & {\textbf{Mean}} & {\textbf{SD}} & {\textbf{CV (\%)}} \\
\midrule
MindistOpp        & 1300.788 & 43.986 & 3.381 \\
MindistScore      & 1285.526 & 43.388 & 3.375 \\
Minscore          & 1263.769 & 39.813 & 3.150 \\
Mindist           & 1259.681 & 41.447 & 3.290 \\
Random            & 569.828  & 42.807 & 7.512 \\
Defeat Heuristic  & 320.049  & 46.737 & 14.603 \\
\bottomrule
\end{tabular}
\caption{\textbf{Rating Summary of Strategies (Custom Elo)}}
\label{tab:rating_summary}
\end{table}

The naive strategies show high coefficient of variation which further indicates inconsistent and highly volatile behavior, reinforcing their inferiority as gameplay approaches.

\section{Discussion}
\label{sec:discussion}

The results presented in Section~\ref{sec:simulation} demonstrate that our proposed modified Elo rating system effectively addresses the fundamental challenges of skill evaluation in luck-influenced, incomplete-information games such as Rummy. This section interprets these findings, discusses their implications, and contextualizes them within the broader landscape of rating systems research.

\textbf{Interpretation of Results}

The simulation study reveals several key insights into the behavior of our proposed rating mechanism. First, the convergence of ratings around 5,000 games indicates that the system successfully stabilizes despite the inherent stochasticity of Rummy. This convergence behavior is critical for practical deployment, as it ensures that player ratings become increasingly reliable indicators of true skill over time, rather than remaining volatile artifacts of recent luck.

The clear hierarchical separation among strategies, particularly the superior performance of hybrid approaches (MindistOpp and MindistScore) over base cases (Mindist and Minscore), validates a core premise of our model: that a rating system incorporating both score-based performance and initial hand quality can effectively discriminate between different levels of strategic sophistication. The fact that these hybrid and basic strategies consistently achieve ratings above 1350, while naive strategies fall well below 1000, demonstrates that our system rewards genuine strategic depth rather than merely reflecting random variation in outcomes.

Moreover, the coefficient of variation (CV) analysis provides important evidence of rating stability. The top-performing strategies exhibit CVs below 4\%, indicating that their ratings, once converged, remain relatively stable despite ongoing matches. In contrast, the Defeat Heuristic strategy shows a CV exceeding 180\%, reflecting its fundamentally inconsistent and self-defeating behavior. This contrast underscores the system's ability to distinguish between stable, skill-driven performance and erratic, luck-dependent outcomes.

\textbf{Advantages Over Traditional Rating Systems}
Our proposed framework offers several methodological advantages over classical rating systems when applied to stochastic environments. To validate this claim, we implemented the traditional Elo system on the same simulation data, with results presented in Appendix~\ref{elo sim}. The comparison reveals that while traditional Elo successfully establishes a similar hierarchical ordering among strategies, our score-based formulation provides additional benefits. Unlike traditional Elo, which relies solely on binary outcomes and thus treats a narrow victory identically to a dominant one, our approach leverages the richer information contained in final scores, rewarding convincing victories while penalizing losses less severely when a player performs respectably despite an unfavorable initial hand.

The incorporation of initial hand quality through the parameter $\beta$ addresses a critical limitation in applying standard rating systems to Rummy. As demonstrated by the parameter tuning procedure (Figure~\ref{fig:paramtune}), optimal calibration of $\beta$ significantly enhances predictive accuracy, with the F1 score reaching 0.7927. This result indicates that explicitly modeling the influence of randomness, rather than treating it as noise, substantially improves the system's ability to infer true skill from observed outcomes.

Compared to more complex systems such as TrueSkill and Glicko-2, our approach maintains computational simplicity while achieving comparable or superior performance in stochastic settings. The deterministic update rule, parameterized by only three constants ($K$, $\alpha$, $\beta$), ensures transparency and interpretability, which are essential for player trust and system adoption in competitive gaming platforms. Furthermore, the zero-sum property inherited from classical Elo ensures that the average rating remains constant, preventing rating inflation or deflation over time.

\textbf{Practical Implications}

While our study focuses specifically on Rummy, the proposed framework is readily adaptable to other games characterized by incomplete information and randomness. The key requirement is the availability of a quantifiable performance metric beyond binary outcomes. For instance, in Poker, chip counts or hand strength distributions could serve analogous roles to MinScore in Rummy. In Bridge, trick counts and bidding accuracy might provide similar informational richness.

For online platforms, our proposed rating system offers several practical benefits. By providing more stable and accurate skill estimates, the system can improve matchmaking quality, ensuring that players are paired with opponents of comparable ability. The system's computational efficiency is also advantageous for large-scale deployment. Unlike TrueSkill or Glicko-2, our model computes rating changes through straightforward arithmetic operations, ensuring low latency and scalability for platforms hosting millions of concurrent matches.

\textbf{Limitations and Considerations}

Despite its strengths, our proposed system has several limitations that warrant acknowledgment. 

\textbf{First}, the simulation study employs deterministic strategies with fixed behaviors, whereas real human players exhibit learning, adaptation, and psychological variability. While our framework is designed to accommodate such dynamics through the sensitivity parameter $K_i$, empirical validation with real player data remains essential to assess its robustness in practice.

\textbf{Second}, the choice of $K$, $\alpha$, and $\beta$ involves subjective calibration decisions. Although we have provided principled justifications for our parameter selections, alternative choices could yield different rating distributions. In particular, the threshold for maximum rating change (set at 50 points in our study) directly determines $K$ and consequently affects convergence speed and long-term stability. Practitioners implementing this system must carefully consider these trade-offs in light of their specific application context.

\textbf{Third}, our current formulation assumes two-player matches. Extending the model to multi-player settings introduces additional complexity, as the zero-sum property must be generalized and benchmark scores must be allocated among more than two players. While conceptually feasible, possibly through pairwise comparisons or rank-based allocations, such extensions require further theoretical development and empirical testing.

\textbf{Finally}, the assumption of a static skill distribution throughout the simulation does not reflect the reality of player improvement over time. Incorporating time-dependent skill evolution, perhaps through decay functions or periodic rating adjustments, could enhance the system's long-term accuracy. The Glicko-2 volatility parameter offers one potential approach to modeling such dynamics, though at the cost of increased complexity.

\section{Conclusion}
\label{sec:conclusion}

This paper has addressed a fundamental challenge in competitive gaming: the accurate evaluation of player skill in environments where outcomes are influenced by both strategic decision-making and random chance. Traditional rating systems, such as Elo, Glicko, and TrueSkill, exhibit significant limitations when applied to incomplete-information, luck-driven games such as Rummy. Our work has developed and empirically validated a modified Elo framework specifically tailored to address these limitations.

The proposed rating system introduces three key innovations: a score-based formulation leveraging richer information from final scores; explicit incorporation of initial hand quality through the parameter $\beta$ to disentangle skill from luck; and a benchmark-based update rule that maintains the zero-sum property while adapting to stochastic gameplay.

Through an extensive simulation study involving 270,000 total games across six strategies, we demonstrated that the system successfully achieves stable convergence after approximately 5,000 games per strategy, effectively discriminates among strategies (with hybrid approaches achieving ratings above 1350 and naive strategies below 500), and exhibits low volatility once converged (CV below 4\% for top strategies). Parameter estimation yielded an F1 score of 0.7927, validating the framework's ability to capture genuine skill differences.

The proposed system offers practical advantages for deployment on online gaming platforms: computational simplicity ensures scalability; interpretability facilitates transparent communication; and stability prevents rating inflation. Beyond Rummy, the framework is readily adaptable to other incomplete-information, stochastic games where quantifiable performance measures exist.

Important limitations include the need for empirical validation with real player data, extension to multi-player settings, and incorporation of time-dependent skill evolution. Future research should address these challenges while exploring adaptive parameters and conducting comparative studies across diverse games.

In conclusion, this work contributes to the growing literature on rating systems for stochastic competitive environments by demonstrating that explicit modeling of randomness and performance quality significantly improves skill inference. The proposed framework offers a principled, computationally efficient, and empirically validated solution with broad potential for application across luck-influenced games. As online gaming continues to expand, rating systems that effectively disentangle skill from chance will become increasingly essential for maintaining competitive integrity, player satisfaction, and engagement.

\section{Appendix}
\subsection{Indian Rummy}
\label{sec:rules}

\begin{itemize}[left=0pt,label={}]
  \item Indian Rummy is a draw-and-discard melding game in which each player is dealt 13 cards and attempts to arrange all cards into valid \textit{sequences} and \textit{sets}; one \textit{pure sequence} (no jokers) is mandatory for a valid declaration. Typical play 2 player version uses one standard 52-card decks together with printed jokers. Below is a brief overview of the rules and scoring system of the rummy game used in our analysis. Details can be found in \cite{rummy}.

  \item \textbf{Core Rules:}
    \begin{enumerate}[left=10pt]
      \item \textit{Deal:} each player receives 13 cards; the remaining stock (closed pile) and one face-up discard form the discard mechanism.
      \item \textit{Turn play:} on a player's turn they must draw one card (from the closed stock or the open discard) and then discard one card to the open pile.
      \item \textit{Valid melds:}
        \begin{itemize}
          \item \textit{Sequence}: three or more consecutive cards of the same suit (e.g., 7$\heart$, 8$\heart$, 9$\heart$).
          \item \textit{Set}: three or four cards of the same rank in different suits (e.g., 8$\club$, 8$\spade$, 8$\diamondc$).
        \end{itemize}
      \item \textit{Declaration:} to declare (win) a player must show at least two sequences, one of which must be a pure sequence (no jokers). Remaining cards must be arranged into valid sets or sequences.
      \item \textit{Jokers \& wildcards}: two printed jokers and one hand-determined wildcards (selected from the exposed card under the stock in many Indian variants) may substitute for missing cards to form impure sequences or sets; a pure sequence cannot contain a joker.
    \end{enumerate}

  \item \textbf{Scoring:}
    \begin{itemize}
      \item \textit{Face cards:} J, Q, K, A = 10 points each.
      \item \textit{Numbered cards:} 2-10 = face value in points (e.g., 5 = 5 points).
      \item \textit{Jokers:} printed jokers and wildcards count as 0 points when used; any un-melded jokers in hand score 0. 
      \item \textit{Round scoring:} after a player declares validly, each opponent totals points from unarranged cards (deadwood). Typical point transfers or losses equal the deadwood counts.
    \end{itemize}
\end{itemize}

For analysis of the game, we next define a few metrics of a given hand, which will help the simulation strategies to take decisions throughout the game. We mainly consider the $MinScore$ and the $MinDist$ metrics of a hand which will in turn, guide play.

\begin{itemize}
    \item \textbf{MinScore:} \label{minscore}  The $MinScore(h, wcj)$ represents the minimum achievable score of a hand $h$ when all cards are optimally arranged into valid melds considering the wild-card joker $wcj$. It quantifies the least possible sum of unmatched card values that can remain after the best possible grouping. A lower $MinScore$ indicates a hand that is closer to a valid declaration, while a higher value reflects a larger number of ungrouped or high-value cards. This metric provides a direct measure of the optimal scoring efficiency of a given hand.
    \item \textbf{MinDist:} The $MinDist(h, wcj)$ metric quantifies the combinatorial distance of a given hand $h$ from the nearest valid configuration considering the wild-card joker $wcj$. It measures how many single-card substitutions are needed to convert the current hand into a valid declaration, thereby capturing the hand's structural proximity to validity. A smaller $MinDist$ indicates a hand that is nearly complete and strategically strong, while a larger value implies greater deviation from a playable configuration. This measure provides an interpretable proxy for the expected number of turns required to achieve a valid declaration.{\color{red}range}

\end{itemize}

Although both metrics measure proximity to a valid declarable hand in distinct ways, each possesses its own advantages and limitations. The $MinDist$ value indicates the number of cards required to complete a valid meld; however, if the needed cards are already in the discard pile, its practical relevance diminishes. On the other hand, the $MinScore$ metric provides a quantitative measure of card arrangement efficiency but lacks a straightforward strategic interpretation. In the following section, we describe the simulation strategies employed in this study, which are formulated based on these two measures.

\subsection{Simulation Strategies}
\label{strategy_descriptions}
\begin{enumerate}
    \item \textbf{Random}: This agent makes completely random decisions at each step, choosing either the deck or pile card, and discarding a random card from its hand.

    \item \textbf{Defeat-Seeking}: This agent identifies existing melds in its hand and intentionally breaks them by discarding the lowest-value card. It picks from the pile or deck based on whether the pile card forms a meld; otherwise, it defaults to the deck. The goal is to maximize the $MinScore$ while disrupting potential meld structures.

    \item \textbf{MinScore-Based}: This agent picks the pile card if doing so improves its $MinScore$ for the best 13 cards by a threshold amount (e.g., 3 points) or completes a valid hand; otherwise, it draws from the deck. It discards the card which reduces $MinScore$ optimally.

    \item \textbf{MinDist-Based}: This agent selects the pile card if it improves its $MinDist$ for the best 13 cards; otherwise, it draws from the deck. It discards the card which reduces $MinDist$ optimally.

    \item \textbf{MinDist + MinScore Hybrid}: This agent prioritizes improving $MinDist$, but when both options yield the same $MinDist$, it chooses the action that results in a better $MinScore$. It follows the same discard criteria as the $MinDist$-based agent.

    \item \textbf{Opponent-Aware MinDist}: This agent extends the $MinDist$-based approach with simple opponent hand modelling. When no improvement in $MinDist$ is possible, it discards cards similar to those recently discarded by the opponent or those likely to form melds with the opponent’s dropped cards (e.g., discarding 4 {$\club$}, 6 {$\club$}, or 5 of another suit if the opponent discards 5 {$\club$}). The agent employs the same drop adherence rule as the $MinDist$-based variant.
\end{enumerate}
\subsection{Simulation Results:}
\label{elo sim}
\begin{figure}[H]
    \centering
    \includegraphics[width=0.5\linewidth]{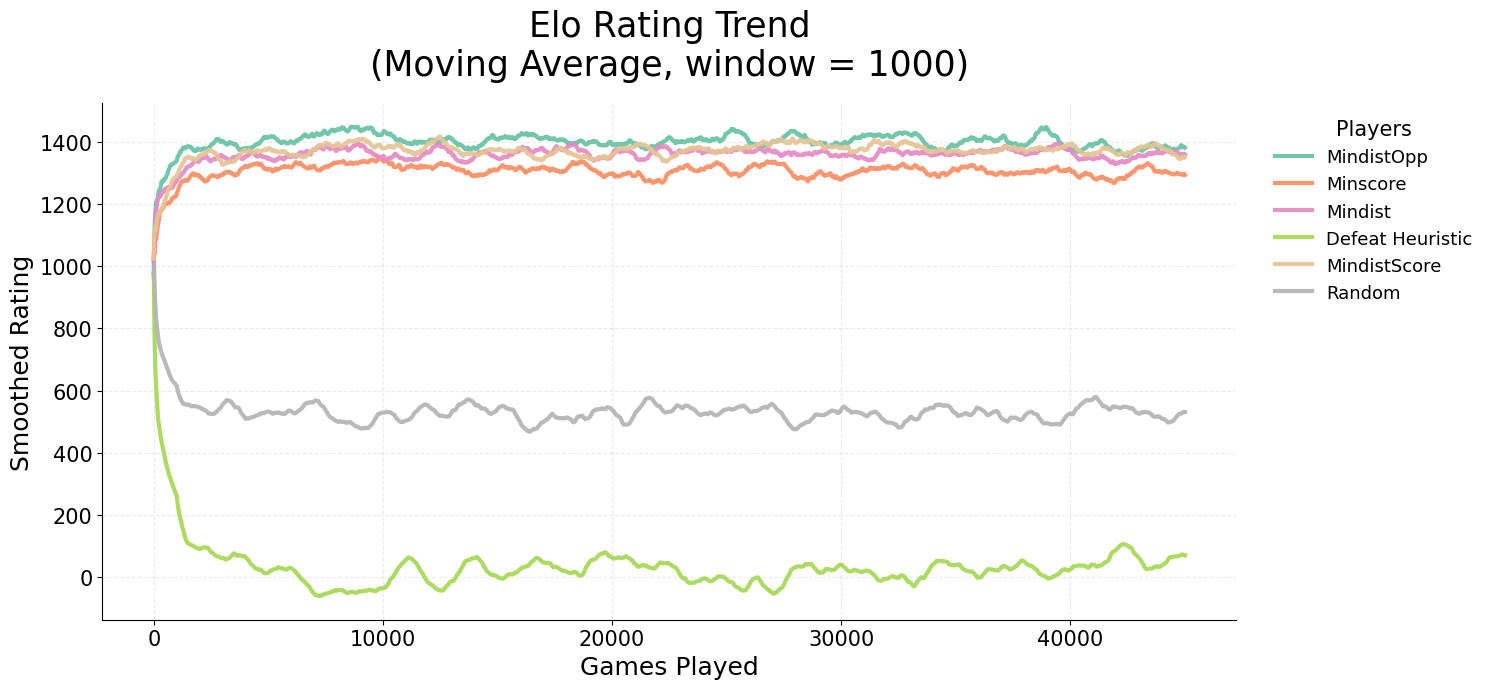}
    \caption{Rating evolution over simulated games (Traditional Elo)}
\end{figure}
\begin{table}[H]
\centering
\renewcommand{\arraystretch}{1.25}
\setlength{\tabcolsep}{8pt}

\begin{tabular}{lrrr}
\toprule
\textbf{Strategy} & {\textbf{Mean}} & {\textbf{SD}} & {\textbf{CV (\%)}} \\
\midrule
MindistOpp        & 1402.204 & 65.425 & 4.666 \\
MindistScore      & 1373.527 & 64.399 & 4.689 \\
Mindist           & 1361.234 & 62.870 & 4.619 \\
Minscore          & 1307.487 & 63.011 & 4.819 \\
Random            & 529.055  & 49.732 & 9.400 \\
Defeat Heuristic  & 26.101   & 67.063 & 256.939 \\
\bottomrule
\end{tabular}
\caption{\textbf{Rating Summary of Strategies (Traditional Elo)}}
\label{tab:elo_rating_summary}
\end{table}
\end{document}